\documentclass[11pt,preprint]{aastex}  
							
\usepackage{geometry}
\usepackage[usenames, dvipsnames]{color}

\usepackage{amsmath,amssymb,graphicx,fancyhdr,natbib, wasysym}

\usepackage[center,tiny]{titlesec}

\begin{document}

\section*{Experimental Constraints On The Fatigue of \\Icy Satellite Lithospheres by Tidal Forces}
\vspace{0.25in}
\centerline{Noah P. Hammond$^{1*}$, Amy C. Barr$^2$, Reid F. Cooper$^1$, Tess E. Caswell$^1$, and Greg Hirth$^1$}

\begin{itemize}
\item[]$^1$Department of Earth, Environmental and Planetary Sciences, Brown University, Providence RI, USA 
\item[]$^2$Planetary Science Institute 1700 East Fort Lowell, Suite 106, Tucson, AZ, USA
\item[]$^*$Now at: Centre for Planetary Sciences, University of Toronto at Scarborough, 1265 Military Trail, Toronto ON, Canada. (noah\_hammond@alumni.brown.edu)
\item[]Accepted for publication in Journal of Geophysical Research - Planets. Copyright 2018 American Geophysical Union. Further reproduction or electronic distribution is not permitted.
\item[]Hammond, N. P. (2018), Experimental Constraints On The Fatigue of Icy Satellite Lithospheres by Tidal Forces , J. Geophys. Res. Planets, in press, doi:10.1002/2017JE005464.
\end{itemize}

\noindent \textbf{Abstract}:  Fatigue can cause materials that undergo cyclic loading to experience brittle failure at much lower stresses than under monotonic loading.
We propose that the lithospheres of icy satellites could become fatigued and thus weakened by cyclical tidal stresses. To test this hypothesis, we performed a series of laboratory experiments to measure the fatigue of water ice at temperatures of $198$ K and $233$ K and at a loading frequency of $1$ Hz. We find that ice is \textit{not} susceptible to fatigue at our experimental conditions and that the brittle failure stress does not decrease with increasing number of loading cycles.  Even though fatigue was not observed at our experimental conditions, colder temperatures, lower loading frequencies, and impurities in the ice shells of icy satellites may increase the likelihood of fatigue crack growth. We also explore other mechanisms that may explain the weak behavior of the lithospheres of some icy satellites.

\baselineskip=20pt   
\parskip=2pt	
\section{Introduction}
Fatigue is an extensively studied phenomenon in engineering, where under certain conditions cyclic loading can cause flaws and microcracks in a material to slowly propagate \citep{paris1963critical, rice1967mechanics, suresh1998fatigue}. As flaws grow, they create stress concentrations in the material that can reduce the brittle failure stress. 

Icy satellites on eccentric orbits about their parent planet experience cyclic loading from tides \citep{burns1986satellites}. We propose that fatigue is a process that could potentially occur in the tidally flexed lithospheres of icy satellites. Cyclic tidal stresses may act upon pre-existing flaws, causing them to slowly propagate, creating stress concentrations near the surface and gradually reducing the brittle failure stress at the surface. The weakening of icy lithospheres by this process would have major implications for the geologic evolution of icy satellites, as the brittle yield stress can control what types of geological process are capable of deforming the surface. A gradual reduction in the strength of the lithosphere through accumulated tidal cycles may allow icy satellites to become more geologically active over time. 

It is unknown whether ice is susceptible to fatigue under tidal loading conditions. Not all materials are susceptible to fatigue, and in materials that are, the rate of weakening can depend on many factors, including temperature, microstructure, chemical environment, and loading frequency (c.f. \citealt{pelloux1969corrosion, stephens1985fatigue}). Laboratory experiments are required to measure the rate of weakening in materials at specific conditions. In this work, we perform fatigue experiments on fresh water ice in order to understand whether fatigue might be occurring in the ice shells of icy satellites. 

Geologic features on icy satellites can provide clues about the strength of their surfaces. Eruptions of water vapor from the south pole of Saturn's satellite Enceladus appear to be regulated by diurnal tidal stresses, suggesting tides are capable of opening surface fractures to expose subsurface water \citep{HurfordNature2007, nimmo2014tidally, porco2014geysers}. On Europa diurnal tidal stresses control the trajectories of long, arch-shaped fractures called ``cycloids'' (see Figure 1), \citep{Hoppa1999,hurford2007cycloidal,rhoden2010constraints}. Solid-state convection in the ice shell may have also deformed the surfaces of several icy satellites (cf., \citealt{parmentier1982,showman2005effects,barr2008,NeillNimmo}). Prominent geologic features on Europa, Ganymede, Enceladus and Miranda are suggested to form as a result of solid-state convection in the ice I shell driving surface deformation (e.g. \citealt{pappalardo1997extensional,Head2002,Prockter2002,pappalardo2004origin,barr2008, hammond2014formation,hammond2014global}).

The magnitude of diurnal tidal stresses on Europa and Enceladus is thought to be $\sim50-100$ kPa \citep{wahr2009modeling, nimmo2014tidally}, and the stresses generated by solid state convection in the ice shell are similarly $10-80$ kPa \citep{showman2005effects,barr2008,NeillNimmo}.
If stresses from both solid state convection and tides can deform the surface, it suggests a surface yield stress for icy satellites of $10-100$ kPa. This is much lower than the brittle yield stress measured in laboratory experiments, typically of order 1 MPa \citep{schulson2001brittle}. There are several possible explanations for this discrepancy. 

\begin{enumerate}
\item The brittle yield stress may be close to $1$ MPa, but stresses from convection or diurnal tides may couple with other sources of stress to overcome the lithospheric yield stress. For example, global volume expansion due to ice shell thickening (e.g. \citealt{nimmo2004stresses}) could create large tensile stresses on which tidal or convective stresses could be superimposed. Additionally, the magnitude of diurnal tidal stresses could be higher than predicted from simple analytical estimates, since tidal stress estimates require modeling the visco-elastic response of the ice shell which is not fully understood (c.f. \citealt{caswell2016convection,beuthe2017enceladus}). 

\item Large scale features tend to fracture at much lower stresses than small scale features because they contain larger flaws (discussed further in section 6.1.2). The lithospheres of icy satellites may be full of flaws that were generated during their formation, or from stresses generated during impact cratering, causing their brittle yield stress to be very low. 

\item The brittle yield stress of icy satellites surfaces may have been reduced due to fatigue from repeated tidal stresses.  Tidal stresses would act upon populations of microcracks and cause them the slowly propagate over billions of tidal cycles. The growth of microcracks throughout the satellite's lithosphere would progressively reduce the brittle yield strength of the surface ice by creating large flaws that could be exploited by weak tidal stresses to cause large-scale fractures. Once a surface has become fatigued, weak tidal and convective stresses may be capable of driving large scale tectonic deformation at the surface. 
\end{enumerate}

Here, we focus on the last hypothesis listed above, and while we explore the possibility of fatigue occurring in icy satellite ice shells, we do not definitively argue that fatigue \textit{is} occurring. We first provide background on the mechanics of fracture and fatigue and previous studies investigating the fatigue of ice. We perform laboratory experiments designed to measure the rate of weakening in water ice at conditions approaching those expected near icy satellite surfaces. We find that fatigue does not occur at our experimental conditions. We conclude by discussing other processes which could weaken the satellites' ice shells to an extent inferred by the observed geology.

\section{Background}
\subsection{The Tensile Failure of Water Ice}
The tensile strength of pure, polycrystalline water ice is influenced by loading history, temperature, porosity and grain size \citep{hawkes1972deformation, schulson1984brittle, schulson1999structure}.  For grain sizes $d=1-2$ mm, or less, the failure stress is related to the grain size by \citep{schulson2001brittle}
\begin{equation}
\sigma_{fail}\approx K d^{-0.5},
\end{equation}
where $K=0.044$ MPa$\text{ m}^{1/2}$ \citep{schulson2001brittle}. For temperatures and grain sizes expected on icy satellites, likely $d\approx 1$ mm \citep{BarrMcKinnon2007}, the brittle yield stress of ice in tension is $1-2$ MPa \citep{litwin2012influence}. The presence of impurities or brines within the ice can also strongly influence the brittle failure stress (c.f \citealt{lee2005mechanics, litwin2012influence}), and these could be important factors that might weaken the surface of icy satellites.

Tensile failure can be described in terms of the Griffith failure criterion and the mode I stress intensity $K_I$ (cf., \citealt{irwin1957analysis}). The stress intensity is a parameter used to estimate the stress environment near a crack tip. The mode I stress intensity for a flaw of length $2a$ under a far field tensile stress $\sigma_{11}$ is
\begin{equation}
K_I=Y \sigma_{11}\sqrt{\pi a}.
\end{equation}
where $Y$ is a geometric factor close to unity. Unstable fracture occurs when the stress intensity reaches a critical value called the fracture toughness $K_{Ic}$ \citep{irwin1957analysis}. 

Materials containing larger flaws fail at lower stresses. For a given stress magnitude, the critical flaw size for brittle failure is 
\begin{equation}
a_{cr}=\frac{2 K_{Ic}^2}{Y^2 \sigma^2\pi},
\end{equation}
where $K_{Ic}\approx0.14$ MPa$\text{ m}^{1/2}$ is the fracture toughness of water ice \citep{nixon1987micromechanical,rist1999experimental,litwin2012influence}.
 
Glaciers and ice sheets tend to fracture at much lower stresses than laboratory ice. Surface crevasses initiate at tensile stresses of $\sigma_t\approx0.1-0.4$ MPa \citep{kehle1964deformation,vaughan1993relating,weiss2004subcritical}. Such low fracture strengths require near-surface flaws $0.1-1.2$ m in length to localize stress. 
If Europa's surface was populated with sharp flaws $\sim5$ m long, it could reduce the surface yield stress to $50$ kPa. In Section \ref{sec:other-mechanisms}, we will discuss how flaws of this size might be initially generated in icy satellite surfaces. 

\citet{weiss2004subcritical} suggest subcritical crack growth as the primary mechanism for the initiation of surface flaws in glaciers. During subcritical crack growth, microcracks slowly propagate at stress intensities below the fracture toughness. Subcritical crack growth has been extensively studied in engineering and geologic materials (cf., \citealt{atkinson1984subcritical, johnson1968sub}), and can occur under either static loading or cyclic loading, when it is usually referred to as fatigue crack growth. Different micro-mechanical mechanisms can contribute to fatigue crack growth. Stress corrosion is the dominant mechanism for subcritical crack growth in most geologic materials. Stress corrosion involves the weakening of molecular bonds near the crack tip by reactive impurities \citep{atkinson1981stress}. 

\subsection{Previous work on the Fatigue of Ice}
Several groups have found that sea ice weakens in response to cyclic loading \citep{haynes1993effect,kerr1996bearing,haskell1996preliminary,bond1997fatigue,langhorne1998break}. These authors generated S-N failure curves which represent the results of fatigue experiments by indicating the number of cycles to failure (N) under a cyclic load of constant magnitude (S) (cf., \citealt{suresh1998fatigue}). These studies find cyclic loading may reduce the brittle yield stress of sea ice by as much as $50\%$ \citep{haynes1993effect,haskell1996preliminary,bond1997fatigue}.  

Additionally, \citet{nixon1987fatigue} found that cyclic loading could reduce the yield stress of fresh water ice by up to 70\%. However, other fatigue studies performed on fresh water ice in the laboratory found that cyclic loading increases the brittle failure stress \citep{cole1990reversed, cole1995cyclic, iliescu2002brittle, iliescu2017strengthening}. \citet{iliescu2002brittle} found that compressive failure stress of S2 columnar ice increased by a factor of 1.5 in response to cyclic loading. More recently, \citet{iliescu2017strengthening} performed cyclic 4-point bending tests on S2 Columnar water ice, and found that the failure stress increased significantly with cyclic loading. They argue that grain boundary sliding during cyclic loading relieved stress concentrations at grain irregularities where cracks initiate.

\citet{weber1997fatigue} have published the only direct measurements of crack growth rates during cyclic loading. They performed cyclic four point bending tests on polycrystalline ice with a pre-cut flaw 5 millimeters in length at the region of maximum tensile stress. The flaw was subjected to stress intensities ranging from $50$ to $90\%$ of the fracture toughness of ice and crack growth was monitored during the experiment using a traveling microscope with a laser illuminating the crack tip. They found that cracks initially grew, but then stopped after growing less than one millimeter, and observed dislocation slip bands ahead of from the crack tip. They suggested that the motion of the dislocations relieved stress ahead of the crack and halted crack growth. \citet{weber1997fatigue} concluded that water ice may not be susceptible to fatigue, however all of their experiments were performed at $T=268$ K, and therefore crack healing and blunting could have strongly inhibited fatigue crack growth. 

We performed experiments to measure the effect of cyclic loading on water ice at temperatures closer to the near surface temperatures found on icy satellites. Icy satellites, however, are likely much more complex in structure than our polycrystalline ice samples. Our samples do not have pre-engineered flaws in them. Thus for fatigue to occur, cracks must either nucleate and then propagate, or cracks must grow from flaws already present at grain boundaries. The question we seek to address through our experiments, however, is could an ice shell that is initially strong ($\sigma_{fail} \approx1$ MPa) be weakened by fatigue.

\section{Laboratory Experiments}
We performed several types of experiments to understand the fatigue of water ice at conditions more relevant to the near surface of icy satellites. Because direct tensile loading of ice is difficult to achieve in the laboratory (see \citealt{cole1990reversed}), we performed indirect tensile loading by the diametrical compression of circular discs (Brazil tests). 

\subsection{Brazil Tests}
Brazil tests are most often used as a simple way to quantify the ultimate tensile strength of materials \citep{hawkes1972deformation}. Our experimental approach is similar to \citet{erarslan2012investigating} who performed cyclic Brazil tests to investigate the fatigue of Brisbane tuff. 
In the Brazil test, circular disks are compressed along their diameter between curved steel plates (see Figure 2) \citep{astm3967}. This loading geometry generates tensile stresses in the center of the sample. Sufficient loading causes fractures to initiate in the center of the sample and propagate parallel to the loading direction. The magnitude of the tensile stresses in the sample center is
\begin{equation}
\sigma_t=\beta \frac{2P}{\pi Dt},
\end{equation}
where $D$ and $t$ are sample diameter and thickness, and $\beta$ is a constant of order $\sim1$ \citep{astm3967}. The stress in the center of the sample when a through going fracture initiates we shall refer to as the ''failure stress''. We also calculated stresses in our ice samples using the finite element package Abaqus \citep{hibbett1998abaqus}. In our numerical model, we replicated our sample loading geometry in two-dimensions, with a refined mesh in areas of high stress (see Supplemental Material).

Brazil tests were conducted in a liquid nitrogen-cooled, ethanol-bath cryostat capable of maintaining sample temperature within $\pm 0.5 ^{\circ}$C \citep{caswell2015constant}. The cryostat is installed on a servo-mechanical Instron 1361 materials testing apparatus. This system is capable of the higher-precision displacement control required to observe fatigue in our experiments. In all tests, prior to applying cyclic loading, the load was increased to $20$ lbs, ($\sim200$ kPa) at a rate of $0.5$ lbs/s and then held constant for $100$ s. This loading procedure allowed a smooth contact to form between the sample and the steel plates and reduced stress concentrations at the contact points. 

Temperatures in icy satellite lithospheres are plausibly less than 150 K, and the tidal stress frequency appropriate for satellites such as Europa and Enceladus, $\omega_{tidal}\sim10^{-5}$ s$^{-1}$.  These conditions are difficult to replicate in the laboratory. One solution to this dilemma is to perform the experiments at a different temperature and stress frequency at which the ice experiences a similar amount of viscous dissipation per cycle as in the natural system,
\begin{equation}
\frac{\omega_{n}\eta_{n}}{\mu}=\frac{\omega_{e}\eta_{e}}{\mu}
\end{equation}
where $\eta$ is the effective viscosity, $\mu$ is the shear modulus, and the subscripts $n$ and $e$ represent the natural and experimental systems. We run experiments at $\omega_e=1$ Hz and $T=198-233$ K, such that viscous deformation is very limited, similar to the conditions expected near the surface of an icy satellite. 

\subsection{Sample Fabrication}
Polycrystalline water ice samples were prepared via the ``standard ice'' method \citep{stern1997grain}. Ice made from deionized water was crushed, sieved to the desired grain size and packed into a mold. The mold was then subjected to vacuum to remove pore gases and subsequently flooded with $0^{\circ}$C, degassed-deionized water. The flooded mold was then allowed to freeze slowly, from the bottom-up, over $24$ hours. This method produced fully dense samples (porosity $<5\%$). Cylindrical samples $2.54$ cm in diameter were cut into cylindrical pucks of thickness $t=1.0 \pm 0.1$ cm using a band saw. Samples had a grain size of $d=0.5-0.8$ mm. Grain size was measured by making surface replicas of ice samples using a liquid Formvar solution, which were photographed with a reflected light microscope (cf., \citealt{klein2010microstructural}). The microstructure of deformed samples were also characterized with this technique. 

To investigate the potential for stress corrosion cracking we performed a suite of fatigue tests on ice samples doped with $50$ ppm H$_2$SO$_4$. These samples were prepared in a manner identical to pure ice samples, except that sulfuric acid was added into the degassed-deionized water that was used to flood the mold. Sulfuric acid naturally concentrates along grain boundaries and triple junctions during freezing (cf., \citealt{baker2003microstructural}). The last sulfuric acid freezes at $198$ K and our ice samples containing sulfuric acid were all tested at $233$ K. Based on the phase diagram of sulfuric acid and water \citep{gable1950phase}, our samples contained $0.1\%$ melt during cyclic loading. We were interested in understanding if partial melts of sulfuric acid, concentrated at grain boundaries, might activate stress corrosion, which involves chemical reactions between fluids within cracks and the crack tip \citep{suresh1998fatigue}.  

\subsection{Fatigue Tests}
Two types of cyclic loading experiments were conducted: constant amplitude tests and increasing amplitude tests. Constant amplitude tests used a constant-amplitude sinusoidal load with a constant mean load. Each of these experiments used a different peak load, varying from $50-90\%$ the load required for monotonic failure, and the minimum load was held at $L_{min}=44.5$ N. The load amplitude $\Delta L$ is defined as the difference between the maximum and minimum load during one loading cycle. For constant load amplitude tests, cyclic loading was applied until the sample fractured or until more than $10^4$ cycles were completed without failure. Some samples that did not fail in cyclic loading were subsequently deformed with a linear load ramp ($44.5$ N/s) to determine the failure stress. For comparison, some linear load tests were also conducted without any cyclic loading. 

In the increasing amplitude tests, a load amplitude $\Delta L$ was applied for $N$ cycles. If failure did not occur, the mean load and $\Delta L$ were increased (keeping $L_{min}$ constant) and $N$ loading cycles were applied again. This process was repeated until the sample failed. At a given temperature, three sets of experiments were performed in which $N = 10$, $100$, or $1000$ cycles were applied before the load was increased. The load amplitude was increased such that the maximum stress increased by $\sim0.1$ MPa. The load and stress amplitude applied during this tests can be found in Table 1. 

Sample failure (crack propagation through the middle of the sample) coincides with a sharp reduction in load and a jump in vertical displacement, both of which were readily detectable in our measurements. To further characterize crack nucleation and growth, we measured acoustic emissions (AEs) and used a small snake-scope camera to monitor the sample surface during the experiments. AEs were measured throughout the experiment (at $10$ kHz) with a contact ultrasonic sensor. The sensor was attached to a Macor$^{TM}$ rod which was in direct contact with the upper steel platen. Figure 3 shows the coincidence between AEs and mechanical data for a monotonic failure test, illustrating that we were able to successfully detect failure. Visual monitoring of the crack also coincided with our measured time of failure, though the camera was not operational for all experiments. 

\section{Experimental Results}

The measured failure stress prior to cyclic loading was $\sigma_{fail}=0.92-1.34$ MPa with an average failure stress of $1.14$ MPa. Considering the maximum grain size in our samples ($d=0.8$ mm), our measured failure stress is similar to but somewhat lower than the predicted tensile yield stress (from equation 1 \citet{schulson2001brittle}) of $1.46$ MPa. \citet{mellor1971measurement} also found that the failure stress of ice during Brazil tests was lower than measured in pure tension. The results of all our experiments are summarized in Table \ref{table:Table1}. 

During the constant amplitude tests, most samples either fractured during the first cycle or they did not fail at all. One sample did not fracture after experiencing up to 75000 cycles with cyclic stress of $\Delta\sigma=0.9$ MPa. One experiment did fail after $100$ cycles, but the cyclic load amplitude was quite high ($1.6$ MPa). AE events were detected in the first few thousand cycles but AE activity decreased with increasing loading cycles. After experiencing over $N=10000$ loading cycles, a linear load ramp was applied to several samples, resulting in a failure stress up to $\sim2$ MPa, significantly higher than failure stress of samples that did not experience cyclic loading. 

During increasing amplitude tests, samples failed as soon as the maximum tensile stress exceeded $\sim1.2$ MPa. Samples that experienced more loading cycles (with $N=1000$ before stress increase), did not fail at lower stresses. The addition of sulfuric acid did not significantly change the tensile failure stress. Experiments at $233$ K generally fail at slightly higher stresses than samples at $198$ K. Colder ice dissipates less energy by viscous creep and more strain energy is available to drive crack growth. Additionally, warmer temperatures may allow flaws and cracks to heal and become blunted, strengthening the ice against brittle failure. This explains why our colder samples fail at somewhat lower loads.  

Figure 4 shows the failure stress of water ice as a function of the number of loading cycles experienced by each sample.  Both the failure stress measured during increasing amplitude tests, as well as the monotonic failure stress measured after constant amplitude tests, are shown. Experiments at all conditions show that cyclic loading does not reduce the tensile failure stress. A slight increase the failure stress is apparent in some cases but overall the failure stress is not strongly affected by the number of loading cycles.  

\section{Discussion}
We find that ice does not weaken when cyclically loaded at $T=193-233$ K with frequency of $1$ Hz. To understand the extent and distribution of microcracking and damage accumulation during our experiments, we have also analyzed our acoustic emission (AE) data.

\subsection{Acoustic Emissions}
Figure 5 shows the mechanical data from our increasing amplitude tests. The stress values plotted are $\sigma_{11}$ values in the center of the sample. AE activity intensified when the maximum tensile stress reached $\sigma=0.78$ MPa and AE events continued to increase in occurrence and amplitude as the cyclic load was increased. The sample failed as soon as the tensile stress reached $1.24$ MPa. Macroscopic crack formation coincided with a load drop, a large AE event and a sharp increase in vertical displacement rate. 

The sharp increase in AE events, when the tensile stress reached $\sim0.8$ MPa, was observed in several experiments and could indicate the onset of crack nucleation in the sample. In water ice under tension with a grain size less than $1-2$ mm, cracks may nucleate at lower stresses before unstably propagating at higher stresses \citep{schulson2001brittle}. The mechanism for crack nucleation in ice is likely dislocation pile-ups at grain boundaries \citep{currier1982tensile,schulson1984brittle,cole1986effect,cole1988crack}, although other mechanisms, such as elastic anisotropy between adjacent grains, may also lead to crack nucleation \citep{sunder1990crack, weiss2000grain}. For dislocations to pile up, dislocations must be generated within the grain (i.e. \citealt{frank1950multiplication}), before they move to the grain boundary. The critical shear stress for initiating crack nucleation by dislocation pile-ups (assuming the slip plane and crack are ideally oriented) is \citep{smith1967crack,cole1988crack},
\begin{equation}
\tau_{cr}=\sqrt{\frac{\pi \gamma_s G }{2(1-\nu)l}},
\end{equation}
where $G=2.3$ GPa is the shear modulus \citep{gammon1983elastic}, $\gamma_s=0.1$ J/m$^2$ is the surface energy of ice \citep{hobbs1974ice}, and $l$ is the characteristic distance over which dislocations pile up. For $l\approx d$, and a grain size of $d=0.5$ mm, we estimate a critical shear stress of $0.8$ MPa for the onset of crack nucleation by dislocation pile-ups, consistent with previous estimates \citep{schulson1984brittle,cole1988crack}. 

Our theoretical calculation of the crack nucleation stress precisely matches the tensile stress in the center of the sample when AE activity increases. However, mysteriously, we do not observe microcracks in our un-failed samples, and we did not observe weakening with cyclic loading. It may be that microcracks formed by this process are similar in length to the grain size and their formation did not increase the largest flaw size available for brittle failure \citep{schulson1984brittle, cole1986effect}.  The paucity of observed microcracks could also be explained by the difficulty of identifying inter-granular fractures (see supplemental material). 

Alternatively, dislocation emission could be responsible for the increased AE activity above $0.8$ MPa, due to reaching the critical stress for generating dislocations. Acoustic emissions associated with the motion of clusters of dislocations have been in detected in ice signal crystals \citep{weiss2001complexity,weiss2003three}. It is uncertain though whether our acoustic sensor would have been capable of detecting dislocation motions, as our sensor was located several inches from the sample. It is our interpretation that the acoustic emmisions increase was due to crack nucleations and that the cracks were likely similar in length to the grain size.   
 
\subsection{Comparison to Other Fatigue Studies}
The bottom panel in Figure 4 shows our S-N failure data compared to the results of previous studies.  We normalize the failure stress measured in each study by the monotonic failure stress measured in each study. Similar to our results, \citet{iliescu2002brittle} and \citet{iliescu2017strengthening} observed that cyclic loading of columnar ice led to an increase in the brittle failure stress, although we do not observe as strong an increase in strength with cyclic loading. In contrast, cantilever beam tests on sea ice show a weakening effect, with the failure stress being reduced by as much as $50\%$ \citep{haynes1993effect,haskell1996preliminary,bond1997fatigue}. 

The presence of salts and brines in the sea ice may promote fatigue and explain the difference in behavior compared to fresh water ice. Stress corrosion requires the presence of a corrosive fluid in the crack (cf., \citealt{atkinson1981stress, atkinson1984subcritical}). Sodium chloride and other salts may be more reactive at crack tips than sulfuric acid, which had little effect in our experiments, making stress corrosion more active in sea ice. Stress corrosion may not occur near the surface of icy satellites where temperatures are low, although it may be feasible deeper in the ice shell where eutectic fluids may be present. In addition to chemical impurities in the material, the presence or absence of atmospheric pressure can affect fatigue crack growth (e.g. \citet{pelloux1969corrosion}). It is unclear what effect, if any, atmospheric pressure might have had on our fatigue experiments, though it would be interesting to perform cyclic loading experiments in vacuum to replicate conditions on icy satellites.

Different loading geometries may also be important in promoting or suppressing fatigue. The only experimental study to show significant fatigue of fresh water ice is \citet{nixon1987fatigue}. Their experiments used a fully reversible twisting load, high frequencies of $2.5$ Hz, and large grain sizes of $d=5-8$ mm. Reversible loading, i.e.,  transitioning between tension and compression, could influence processes occurring near the crack tip.   Fully reversible loading also prevents permanent deformation of the sample, which can affect the stress distribution even if the load is constant (see Supplemental Material). Under tidal loading conditions the stress alternates between tensile, shear and compressive \citep{helfenstein1983patterns}, similar to the reverse bending test performed by \citet{nixon1987fatigue}. Grain size may also be important because large grain sizes can suppress grain size sensitive creep, and make dislocation creep the dominant creep mechanism. \citet{iliescu2017strengthening} suggested that grain boundary sliding during cyclic loading can lead to strengthening and it has been argued that dislocation cross slip (the rate limiting step in dislocation creep) may be essential to promoting fatigue crack growth \citep{cole1990reversed}. 

\section{Implications for Icy Satellites}
We do not observe fatigue in our experiments, however, we have so far only explored a small range of experimental conditions. Fatigue experiments on metals and ceramics show that slight changes in material properties or test conditions can produce dramatic differences in fatigue behavior \citep{suresh1998fatigue}. Colder temperatures, lower loading frequencies, and the presence of corrosive fluids, all of which would be expected in planetary ice shells \citep{collins2009tectonics}, may promote fatigue. It is also possible that the conditions that promote fatigue in ice are not present on icy satellites, and that there is another explanation for the apparent weak behavior of icy satellites. Other potential weakening mechanisms include: 1) the generation of flaws in the ice shell from impact cratering, 2) the superposition of tidal stresses with other larger sources of tensile stress, 3) the inherent weakness of larger scale structures due to inhomogeneities. 

\subsection{Other Weakening Mechanisms} \label{sec:other-mechanisms}
\subsubsection{Superposing Stress Fields}
Several other mechanisms may be capable of generating large stresses on icy satellites that convective and diurnal tidal stresses could combine with to fracture the ice shell. These processes include global surface expansion, polar wander, obliquity tides and non-synchronous rotation of the ice shell with respect to the core \citep{collins2009tectonics}. 

There is evidence that multiple stress mechanisms may have combined to fracture the surfaces of various icy satellites. On Ganymede, weak convective stresses were likely important in the formation of grooved terrain \citep{hammond2014formation}, but it is possible that grooved terrain formed due to convective stresses combining with stresses from global surface expansion \citep{ShowmanMalhotra1997}. On Europa, stresses from diurnal tides are thought to be the main driver of cycloid formation \citep{Hoppa1999}, however stresses from non-synchronous rotation could combine with diurnal tides and increase the magnitude of the cyclic stresses \citep{hurford2007cycloidal, rhoden2010constraints, wahr2009modeling}.  It remains a strong possibility that weak tidal and convective stresses are only capable of deforming the surface because they are combining with other larger sources of stress. In this case, fatigue crack growth from tidal stresses would not be necessary to explain how tides could fracture the surface. 

\subsubsection{Large-Scale Inhomogeneities}
On Earth, large scale features naturally form fractures at much lower stresses. For instance, \citet{dempsey1999scale} studied the failure strength of sea ice beams over a large size range and found that $80$ m long beams failed at tensile stresses as low as $40$ kPa. One argument to explain this behavior is that larger scale features contain larger flaws that localize stress. Scale-dependent failure models often use a Weibull flaw distribution, where the failure strength for a volume of material $V$,
\begin{equation}
\sigma_{fail}\sim V^{-1/12}.
\end{equation}
This relationship was empirically derived by \citet{parsons1992influence} who measured the failure strength of sea ice and icebergs over a large range in sizes. It is unclear that this relationship is applicable at the global scale but it would imply that icy satellite lithospheres should behave 10 to 100 times weaker than small ice samples tested in the laboratory.

On Europa, diurnal tidal stresses of $50$ kPa in magnitude appear to cause widespread tensile failure. If Europa's lithosphere was full of sharp flaws $\sim 5$ meters in length, Europa's low brittle yield stress could be explained. Based on the size effect discussed above, it may be likely that flaws of this size would exist at the global scale. Possible mechanisms for the initial generation of such flaws include the freezing of brine pockets, thermal expansion stresses, or damage from impact cratering. 

\subsubsection{Damage from Impact Cratering}

Impact cratering can generate extensive fracturing on planetary surfaces. Most of the damage from an impact is localized near the surface in a thin layer of highly porous debris called ``regolith.'' On Europa, the estimated impact regolith thickness is $0.1-1$ m \citep{moore2009surface}. However, impact cratering may cause fracturing at much greater depths due to impact shock waves propagating into the subsurface. Viscous relaxation or thermal contraction after the impact can also cause stress and damage in the lithosphere \citep{bratt1985evolution}. A deeper fractured zone called the ``mega-regolith'' layer can form as a result of impact cratering. The mega-regolith on Earth's Moon is $5-15$ km thick, and has an estimated $10-20\%$ porosity \citep{besserer2014grail,soderblom2015fractured}. 

Impacts into ice show that fractures form in response to the passage of impact shock waves \citep{arakawa1995direct}, and the depth of fracturing can be predicted by comparing the shock wave pressure to the brittle failure stress \citep{ahrens2002depth}. Impact shock waves generate compressive radial stresses that result in relatively tensional concentric stresses, leading to the formation of radial fractures deep within the target \citep{ai2004dynamic}. Shock wave pressure as a function of radial distance $r$ from the impact point can be approximated as \citep{MeloshBook}
\begin{equation}
P(r)=P_0(r/r_i)^{-n},
\end{equation}
where $r_i$ is the radius of the impactor, $P_0$ is the peak pressure near the impact point, and $n$ is the decay exponent. The peak contact pressure during a high velocity impact can be approximated as $P_0\approx \rho_i v^2$ where $\rho_i$ is the density of the impactor and $v$ is the impact velocity \citep{MeloshBook}. For ice at $T=150$ K, the shock wave decay exponent $n=1.5$ for low pressures and $n\approx4$ for pressures above $300$ MPa, due to energy absorbed by high pressure phase transitions in ice \citep{kraus2011impacts}. Further away from the crater, the shock pressures decay elastically, with a decay exponent of $n=2$. The concentric stresses associated with the passage of the shock wave are $\sigma_{\theta\theta} = -1/3 P(r)$ \citep{ai2004dynamic}.

Figure 6 shows the concentric stresses generated by a comet impacting at a velocity of $v=5$ km/s with an impactor diameter of $r_i=300$ m. On Europa, such an impact would result in a crater with a diameter $D\approx10$ km. Stresses caused by the shock wave would be capable of causing tensional failure to a depth of at least 4 km. 

Our analysis suggests that even relatively small and low-velocity impactors could generate a significant amount of subsurface damage. Although stresses generated during the formation of a 10 km diameter impact crater could cause significant damage, not many larger craters exist on Europa's surface. Therefore, recent impact cratering is unlikely to generate widespread and deep lithospheric damage on Europa. However, its possible that damage from an earlier more intense period of impact cratering could have remained in the lithosphere. As discussed below, cracks and damage may not heal at the cold surface temperatures of icy satellites. 

\subsection{Crack Healing}
Several processes could contribute to the removal of cracks in an ice shell including elastic compaction, bulk viscous strains from convective flow, the refreezing of melt in the crack and surface energy reduction accommodated by diffusion. Elastic compaction is not very efficient at closing pore space on icy satellites due to the low gravity and low pressure gradients. \citet{durham2005cold} showed that ice could sustain a porosity of over $10\%$ at pressures up to $150$ MPa. Viscous strains from convection would likely efficiently remove cracks. The questions then becomes: how deep in the lithosphere of the ice shell could cracks be present and how quickly would they heal due to diffusion.

There are surprisingly few studies of crack healing in ice. \citet{barrette2002healed} studied crack healing in icebergs, but these cracks began healing by being refilled with melt that refroze. This type of process may only occur at the base of the ice shell, but could also occur below chaos terrain on Europa where shallow melt is expected \citep{Schmidt2011}. \citet{colbeck1986theory} discussed a theory for microcrack healing in ice driven by vapor diffusion, but we focus on crack healing driven by crack tip radius curvature limited by volume diffusion. We know of one study that observed the healing rates of microcracks in ice \citep{cole1986effect} and although their data is limited our crack healing model is consistent with their observations (see Supplemental Material).

For simplicity, imagine a straight crack with an initial length $a$, and a constant crack tip radius $r$. The driving force for crack retreat is proportional to the curvature of the crack tip radius, $F_b=\gamma/r$, where $\gamma=0.1$ J/m$^2$ is the surface energy of ice \citep{hobbs1974ice}. The velocity of the crack boundary is $V_b=F_b M_b$ (e.g., \citealt{evans2001few}), where $M_b$ is the mobility of the crack tip. 
In the case where crack healing is accommodated by grain boundary diffusion, the mobility, 
\begin{equation}
M_b=\frac{\Omega D_{gb}}{\delta k_{bz} T},
\end{equation}
%
where $\Omega$ is molar volume, $k_{bz}$ is Boltzmann's constant, $\delta$ is the width of the crack boundary and $D_{gb}$ is the grain boundary diffusion coefficient. From Ramzier (1967), $D_{gb}=D_o \exp((-Q)/RT)$ and $D_0$ is constrained from experiments by \citet{GoldsbyKohlstedt2001}, and $R=8.314$ J/mol-K is the gas constant. The grain boundary mobility can also be calculated from parabolic grain growth experiments (e.g., \citealt{evans2001few}). 

Assuming the crack tip radius remains constant with time, the crack healing rate can be written as:
\begin{equation}
\frac{da}{dt}=\frac{\gamma}{r}\frac{\Omega D_o}{\delta k_{bz} T } \exp(-Q/RT)
\end{equation}

From this equation we predict that the rate of crack healing during our fatigue experiments is negligibly small. Over the course of our longest experiment ($8\times10^4$ s) and for a crack tip radius of $r=0.01$ mm, we predict a microcrack would have only shortened due to healing by 0.02 mm. Our calculated rates are consistent with observations by \citep{cole1986effect}. Very sharp cracks however, with $r<1 \mu$m, could have annealed during our experiments.  

Figure 7 shows how long it would take to heal a crack $a=1$ m long for a given temperature. Our analysis suggests that in order to completely heal a $1$ m long crack after $1$ Myr, for a crack tip radius of $r=0.01-0.1$ mm, temperatures $T>170-200$ K would be required. In the near surface of icy satellites where $T<170$ K, cracks could remain present for long time periods without healing due to diffusion. Below this depth, there could still be fractures and microcracks, but they could potentially heal over geologic timescales. We predict that cracks formed in an ice shell billions of years ago would not heal at the near-surface temperatures of the outer solar system. Thus any damage produced during ice shell formation or impact bombardment may still be present, unless the near-surface ice has been remelted or recycled into the interior. 

\section{Conclusions}
We performed cyclic loading experiments on polycrystalline water ice, using a Brazil test geometry, at temperatures of $T=198$ K and $T=233$ K, a loading frequency of $\omega=1$ Hz, and with either a gradually increasing cyclic load amplitude or constant cyclic load amplitude. In all cases we find that ice is not susceptible to fatigue at our experimental conditions. An increase in acoustic emissions activity when tensile stress in the center of the samples reached $\sigma=0.8$ MPa is interpreted as the onset of crack nucleation. The lack of weakening in our samples suggests that microcracks did not grow under the action of cyclic loading. Fatigue in ice satellites, however, cannot be ruled out. More experiments are required to better understand subcritical crack growth and damage accumulation at temperatures, frequencies and chemical environments that more closely approach the conditions in icy satellite lithospheres. 

There are several processes, other than fatigue by tidal forces, that can influence the strength of icy satellite lithospheres, including damage from impact cratering, naturally large flaws due to the large scale of icy satellite surfaces and the coupling of weak stresses with other sources of stress. These effects may work in tandem on different icy satellites to control the effective strength of the surface, and the unique geologic history of each icy satellites may influence which of these processes dominates the strength of the lithosphere. We find that the formation of a 10 km diameter impact crater on Europa could cause fracturing to a depth of up to 4 km. Our calculated crack healing rates in the ice shell show that in the lithosphere where $T<180$ K, large cracks could remain present for millions to billions of years. This suggests that the lithosphere of Europa could support a high density of fractures down to depths of $1-5$ km, depending on the thermal gradient and its resurfacing history. 

Fatigue in water ice remains a poorly understood yet highly important subject, because cyclic loading of ice occurs not only on icy moons, but on ice shelves on Earth as well. Future experiments on ice fatigue with more detailed observations of microcrack behavior will enable us to better understand and predict icy satellite tectonics and ice rift propagation. 

\section{Acknowledgements}  Author Hammond was supported by NASA NESSF grant NNX13AN99H. Author Barr acknowledges support from NASA OPR NNX09AP30G.  Authors Cooper and Caswell acknowledge the support of NASA Solar System Workings grant NNX16AQ14G. All data necessary to reproduce our results are included in the manuscript text and in Table \ref{table:Table1}.

\clearpage
\begin{table}
\caption{Data showing the total number of loading cycles experienced versus the failure stress for ice fatigue experiments at 198 K and 233 K, a loading frequency of 1 Hz and a grain size of $d=0.5$ mm. Each row describes an experiment preformed on one ice sample. The results of several test types are shown: 1) Load Ramp: a monotonic failure test where the load was increased by 10 lbs/s until failure, 2) Constant $\Delta\sigma$: cyclic loading at a constant $\Delta L$, 3)Load ramp after constant $\Delta\sigma$, and 4) Increasing $\Delta\sigma$: $\Delta L$ was increased after $N$ number of cycles. $\Delta\sigma$ is the stress amplitude in the center of the sample for a given load amplitude $\Delta L$. $^*$In these experiments the ice samples contained 50 ppm H$_2$SO$_4$. } \label{fatiguetable} \label{table:Table1}
\begin{tabular}{l l l l l}
\hline
\small{\textbf{Temp (K)}} & \small{\textbf{Cyclic Stress (MPa)}} & \small{\textbf{Test Type}} & \small{\textbf{Total No. of cycles}} & \small{\textbf{Failure Stress}}\\
\hline
233 & - & Load Ramp & $<1$ & 1.12\\
233 & - & Load Ramp & $<1$ & 1.40\\
233 & - & Load Ramp & $<1$ & 1.10\\
233* & - & Load Ramp & $<1$ & 1.10\\
198 & - & Load Ramp & $<1$ & 0.92\\
233 & 0.9 & Constant $\Delta\sigma$ & 104 & 0.92\\
233 & 1.0 & Load Ramp after Constant $\Delta\sigma$ & 10000 & 1.675\\
233 & 1.0 & Load Ramp after Constant $\Delta\sigma$ & 19000 & 2.28\\
233 & 0.9 & Load Ramp after Constant $\Delta\sigma$ & 75000 & 1.88\\
233 & \textit{0.6, 0.7, 0.8 ...}1.7 & Increase $\Delta\sigma$ after N=10 & 101 & 1.70\\
233 & \textit{0.6, 0.7, 0.8 ...}1.5 & Increase $\Delta\sigma$ after N=100 & 902 & 1.53\\
233 & \textit{0.6, 0.7, 0.8 ...}1.6  & Increase $\Delta\sigma$ after N=1000 & 11002 & 1.65\\
233* & \textit{0.4, 0.55, 0.7 ...}1.25 & Increase $\Delta\sigma$ after N=10 & 69 & 1.26\\
233* & \textit{0.37, 0.5, 0.63 ...}1.37 & Increase $\Delta\sigma$ after N=100 & 810 & 1.37\\
233* &\textit{0.42, 0.56, 0.7 ...}1.4  & Increase $\Delta\sigma$ after N=1000 & 7726 & 1.40\\
198 & \textit{0.75, 0.87, 1 ...}1.25 & Increase $\Delta\sigma$ after N=10 & 45 & 1.25\\
198 & \textit{0.5, 0.6, 0.7 ...}1.6 & Increase $\Delta\sigma$ after N=100 & 727 & 1.12\\
198 & \textit{0.7, 0.8, 0.9 ...}1.5 & Increase $\Delta\sigma$ after N=1000 & 7500 & 1.46\\
\hline
\end{tabular}
\end{table}

\clearpage
\begin{figure}
\centering\includegraphics[width=30pc]{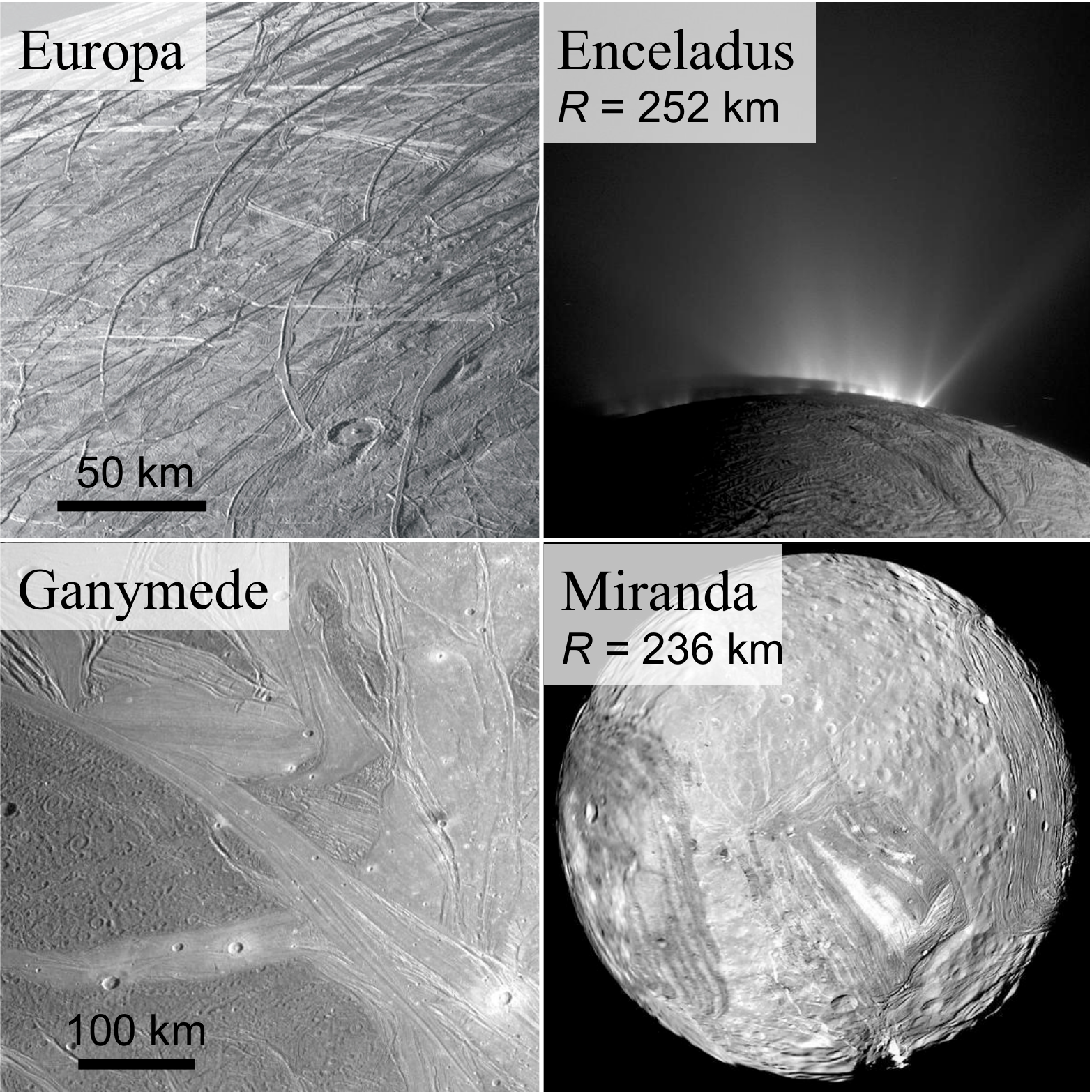}
 \caption{Examples of geological features on icy satellites that suggest their surfaces might be weak. (top left) Arch shaped fractured (cycloids) on Europa form in response to tidal stresses $\sim50$ kPa in magnitude. (top right) Eruptions of water vapor plumes on Enceladus are regulated by tidal stresses of $<100$ kPa. Both light grooved terrain on Ganymede (bottom left) and coronae on Miranda (bottom right) are suggested to form by convection driven resurfacing, which would required the brittle yield stress of the surface to be less than $\sim50$ kPa. Image credit: NASA \textit{Galileo} SSI image numbers 15E0096, G8G0001, NASA Planetary Photojournal \textit{Cassini} ISS image PIA17184 and \textit{Voyager} SSI mosaic by Ted Styrk. \label{fig:Figure1}}
 \end{figure}

\clearpage
\begin{figure}
\centering\includegraphics[width=30pc]{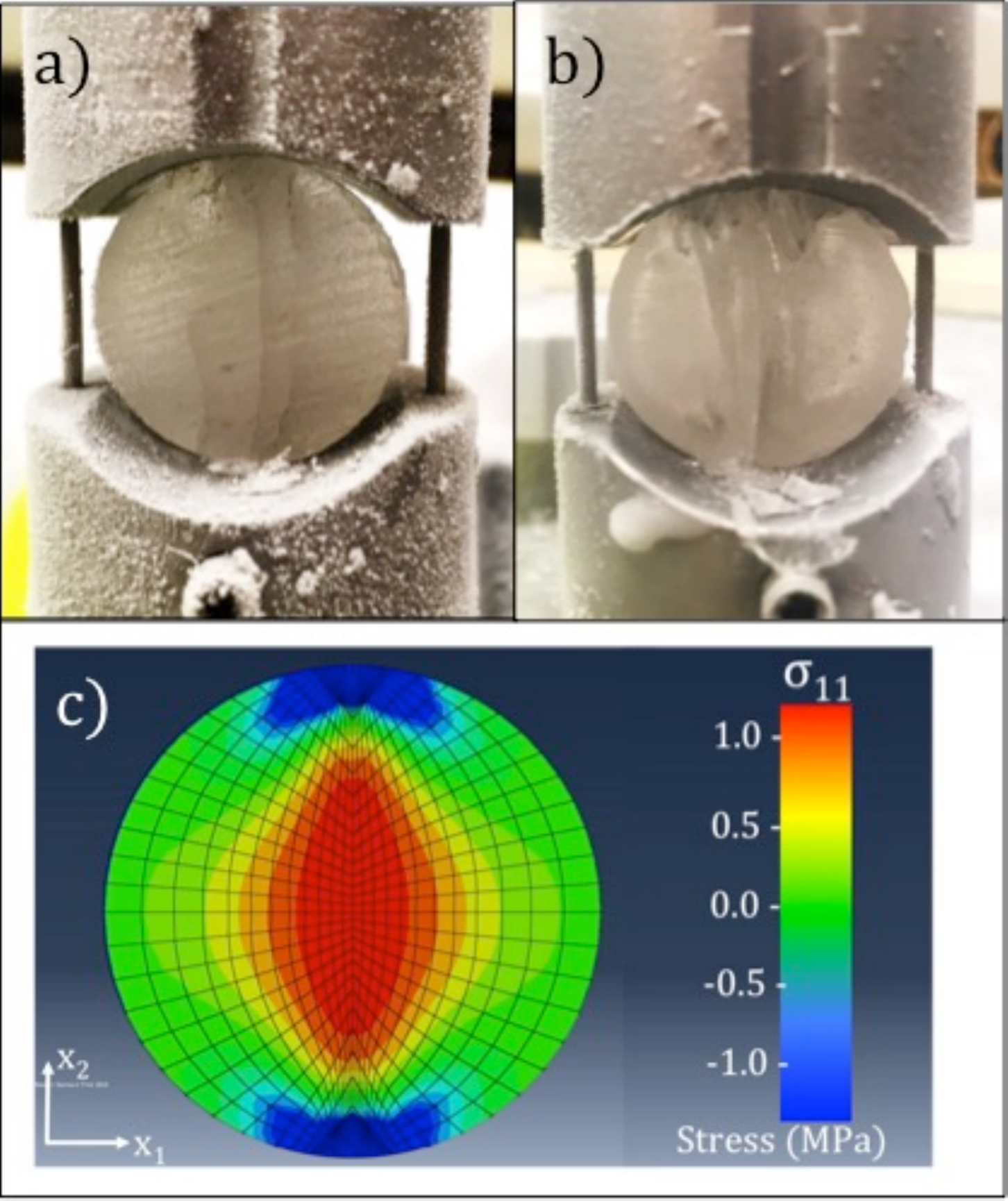}
\caption{a,b) Example of failed ice samples after cyclic Brazil test. Ice samples are circular discs with a diameter of $2.54$ cm and a thickness of $1.0$ cm. Curved steel platens are used as contact pieces to alleviate stress concentrations at loading points. Failed samples show vertical fractures that propagate through the center of the sample. c) Numerical calculations of the $\sigma_{11}$ stress during a Brazil test. Red colors show tensile stresses and blue show compressive stresses. Simulations were performed with the numerical model Abaqus \citep{hibbett1998abaqus} and used a fixed boundary condition at the base of the disc and a fixed load of  $100$ lbs at the top of the disc. See Supplementary Material for more details.  \label{fig:Figure2}}
\end{figure} 

\clearpage
\begin{figure}
\centering\includegraphics[width=25pc]{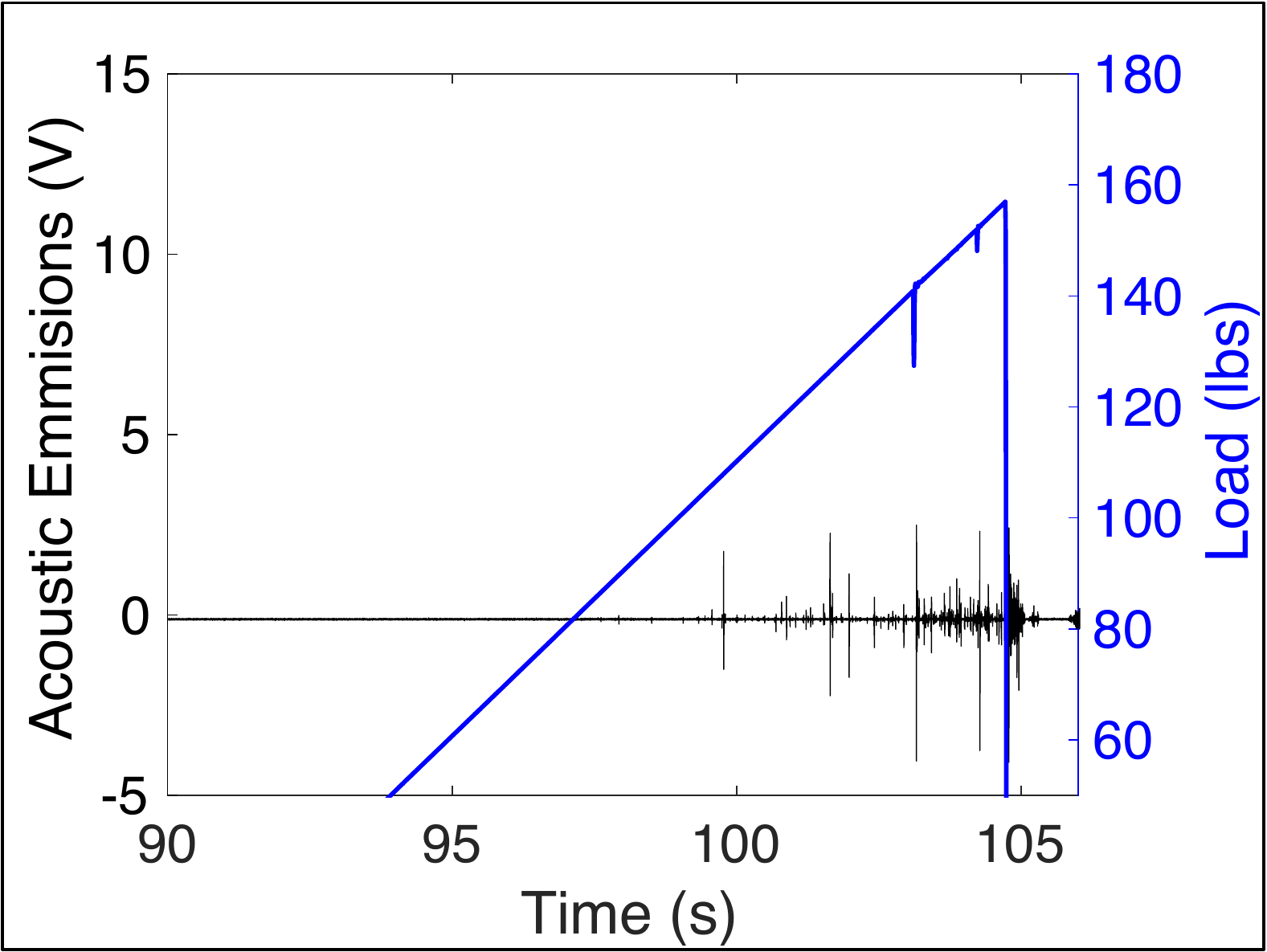}
\caption{Acoustic emissions data measured during a Brazil test for monotonic failure. Small events prior to failure are interpreted as crack nucleation events. Initial load failure occurs at $138$ pounds. This corresponds to a tensile stress in the sample center of 1.4 MPa (using equation 4 and a sample thickness of 1.1 centimeters. The load drop corresponds with a larger acoustic events. The failure event coincided with the appearance of a vertical fracture in the sample. \label{fig:Figure3}}
\end{figure} 
 
\clearpage
\begin{figure}
\centering\includegraphics[width=25pc]{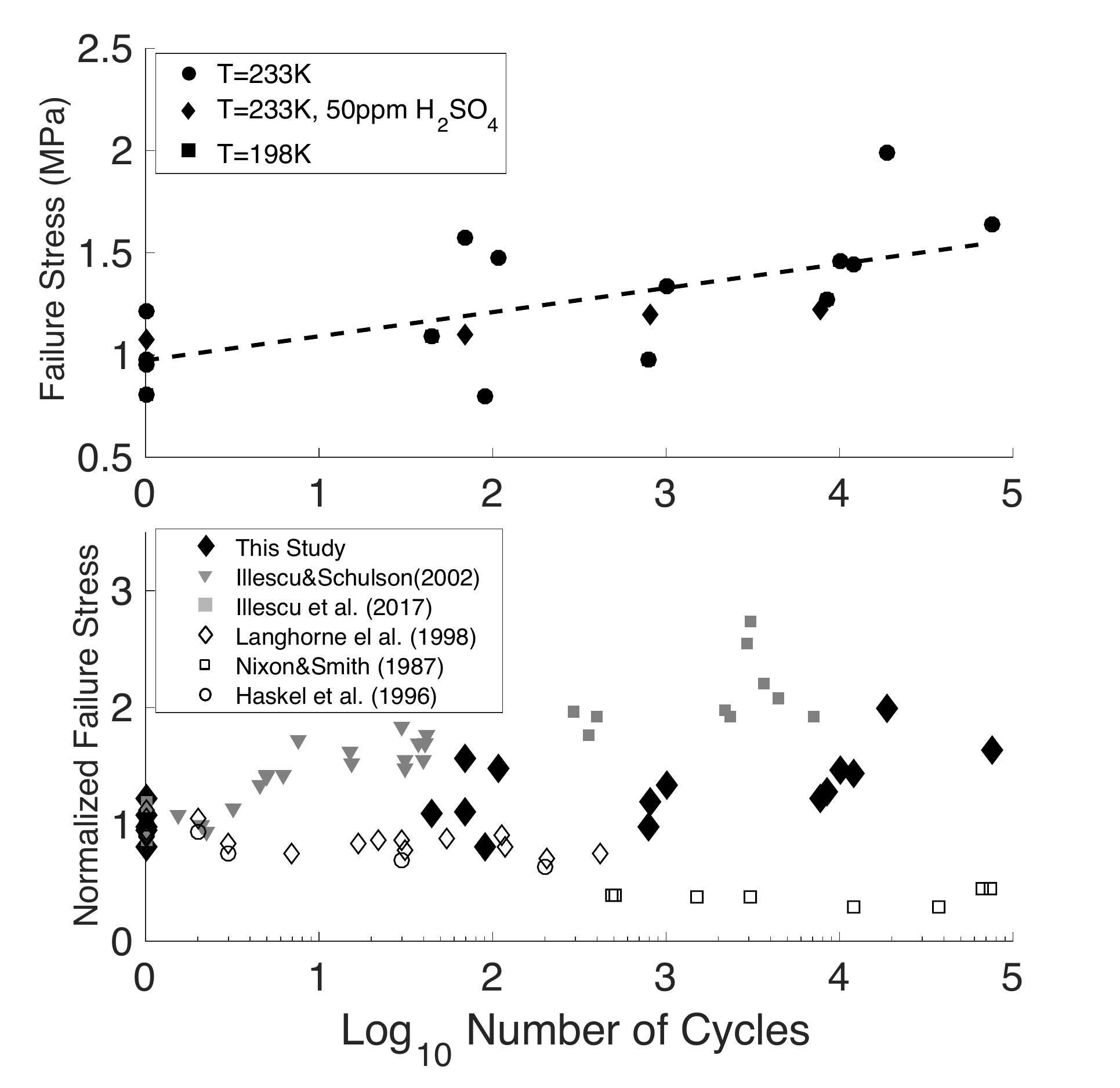}
\caption{(top) Failure stress versus the number of load cycles experienced. Data from both types of fatigue tests are included, (gradually increasing cyclic load and the monotonic failure after constant cyclic load tests). Circles and squares show experiments performed on pure polycrystalline ice at $233$ K and $198$ K, respectively, and diamonds show experiments performed on ice doped with 50 ppm H$_2$SO$_4$. A dashed line is a fit to the data of the form $\sigma_f=0.974+0.118\log_{10}(N)$, which can explain $42\%$ of the variance. (bottom) Comparison of our data to previous studies of ice fatigue. The failure stress was normalized by the monotonic failure stress measured in each study. Black diamonds show all of our data. In situ sea ice fatigue data are represented by open diamonds \citep{langhorne1998break}, and open circles, \citep{haskell1996preliminary}. Experiments on fresh water ice are shown by open squares (\citet{nixon1987fatigue}), gray triangles (\citet{iliescu2002brittle}) and gray squares (\citet{iliescu2017strengthening}). \label{fig:Figure4}}
\end{figure}

\clearpage
\begin{figure}
\centering\includegraphics[width=30pc]{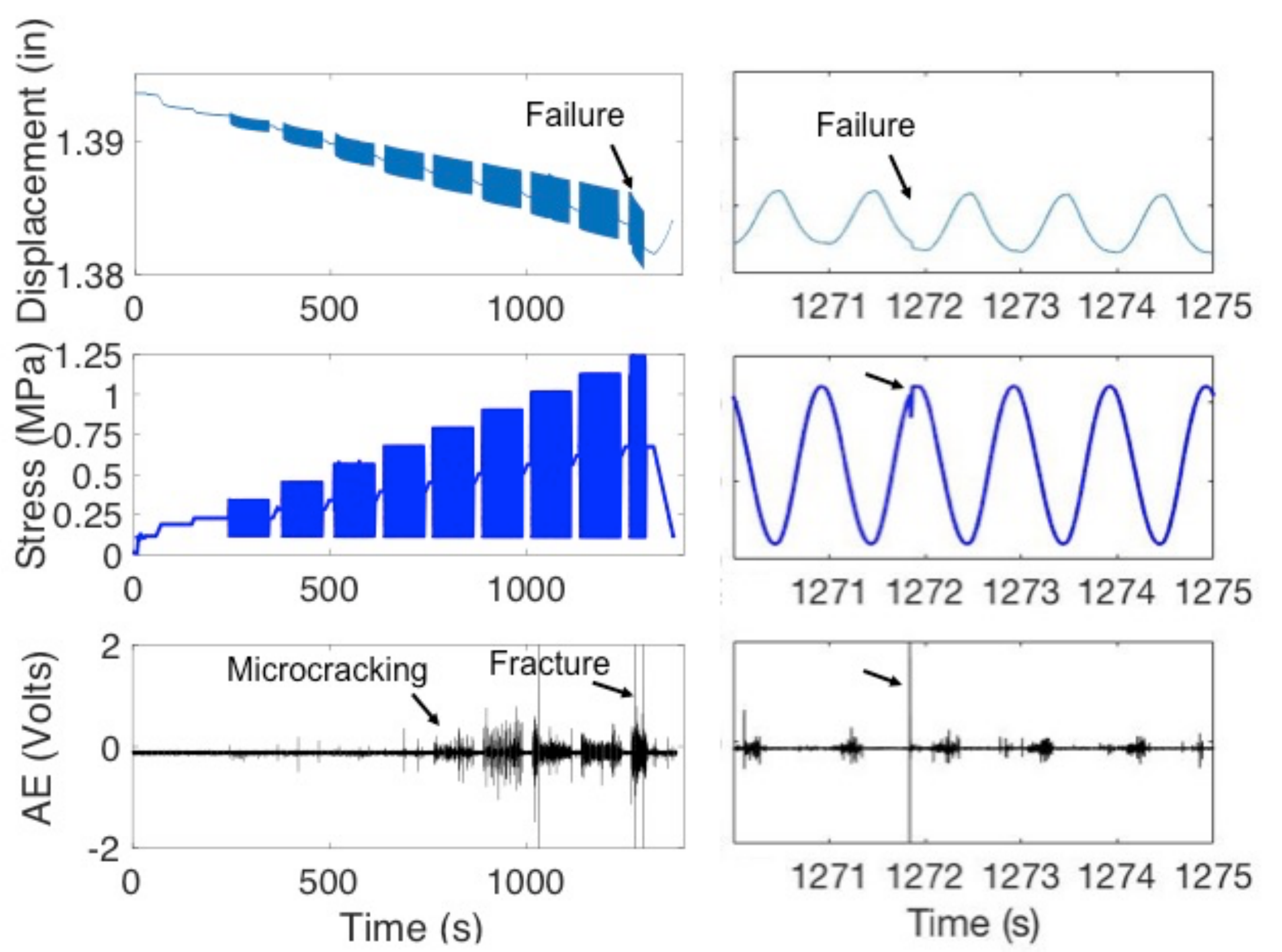}
\caption[Mechanical data during increasing amplitude test]{Mechanical data recorded during increasing amplitude fatigue tests on ice, showing vertical displacement (top), load (middle) and acoustic emissions (bottom) with time on the x-axis. Significant acoustic emission events begin at 750 seconds, interpreted as microcracking in response to cyclic loading. The sample fails at 1271.8 seconds, with a transition to tertiary creep, load drop, and a large AE event. This failure event is interpreted as the formation of a vertical fracture through the middle of the sample\label{fig:Figure5}.}
\end{figure}

\clearpage
\begin{figure}
\centering\includegraphics[width=30pc]{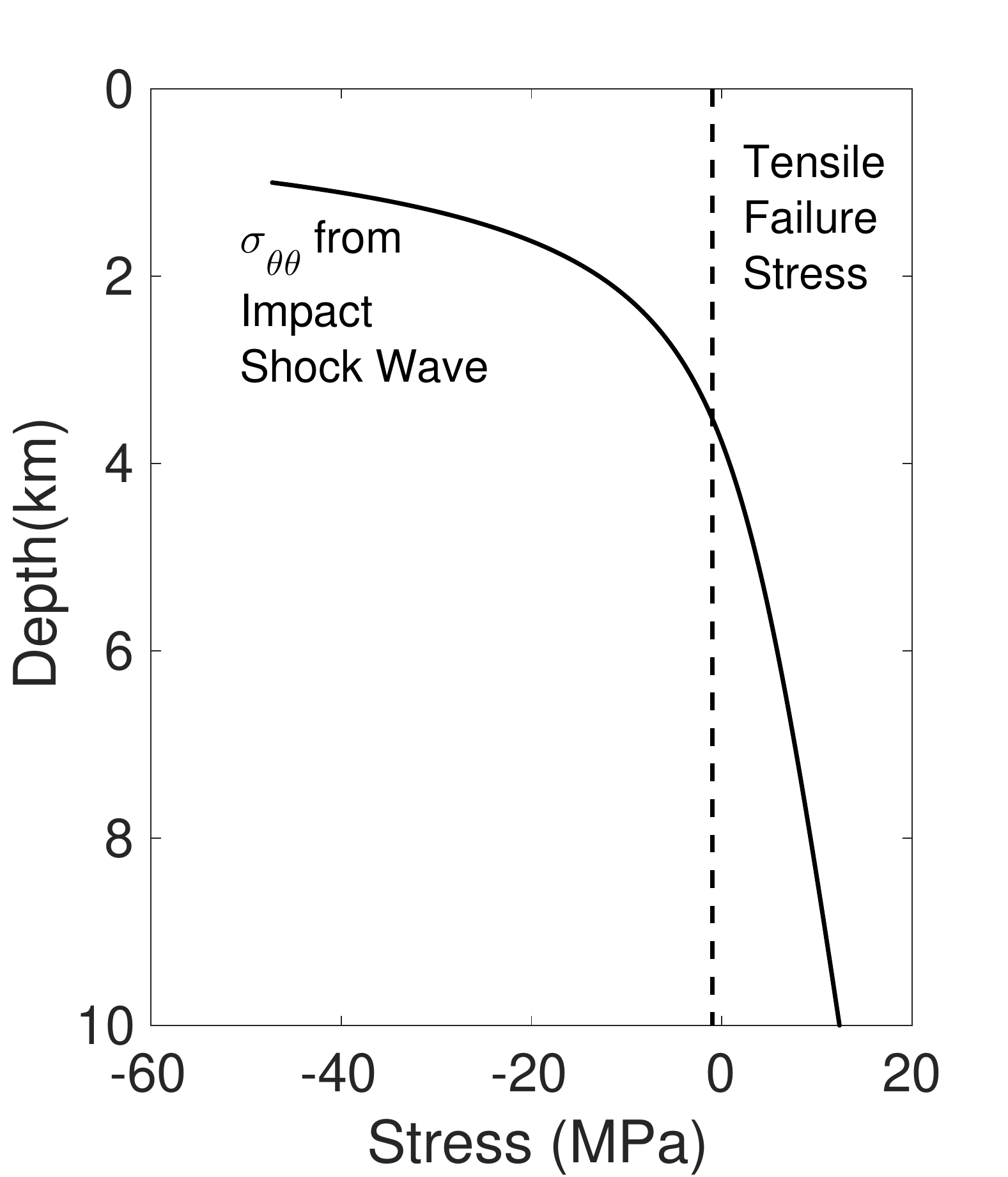}
\caption[Magnitude of impact shock waves with depth in the ice shell compared to yield stress]{ Impact-generated stresses with depth due to the formation of a 10 km diameter crater on Europa. Solid line shows how the concentric stresses from the impact shock waves decay with depth. Overburden stresses also reduce the magnitude of tensile stresses (depicted as negative). Dashed line shows the stress required for tensile failure. \label{fig:Figure6}}
\end{figure} 

\clearpage
\begin{figure}
\centering\includegraphics[width=30pc]{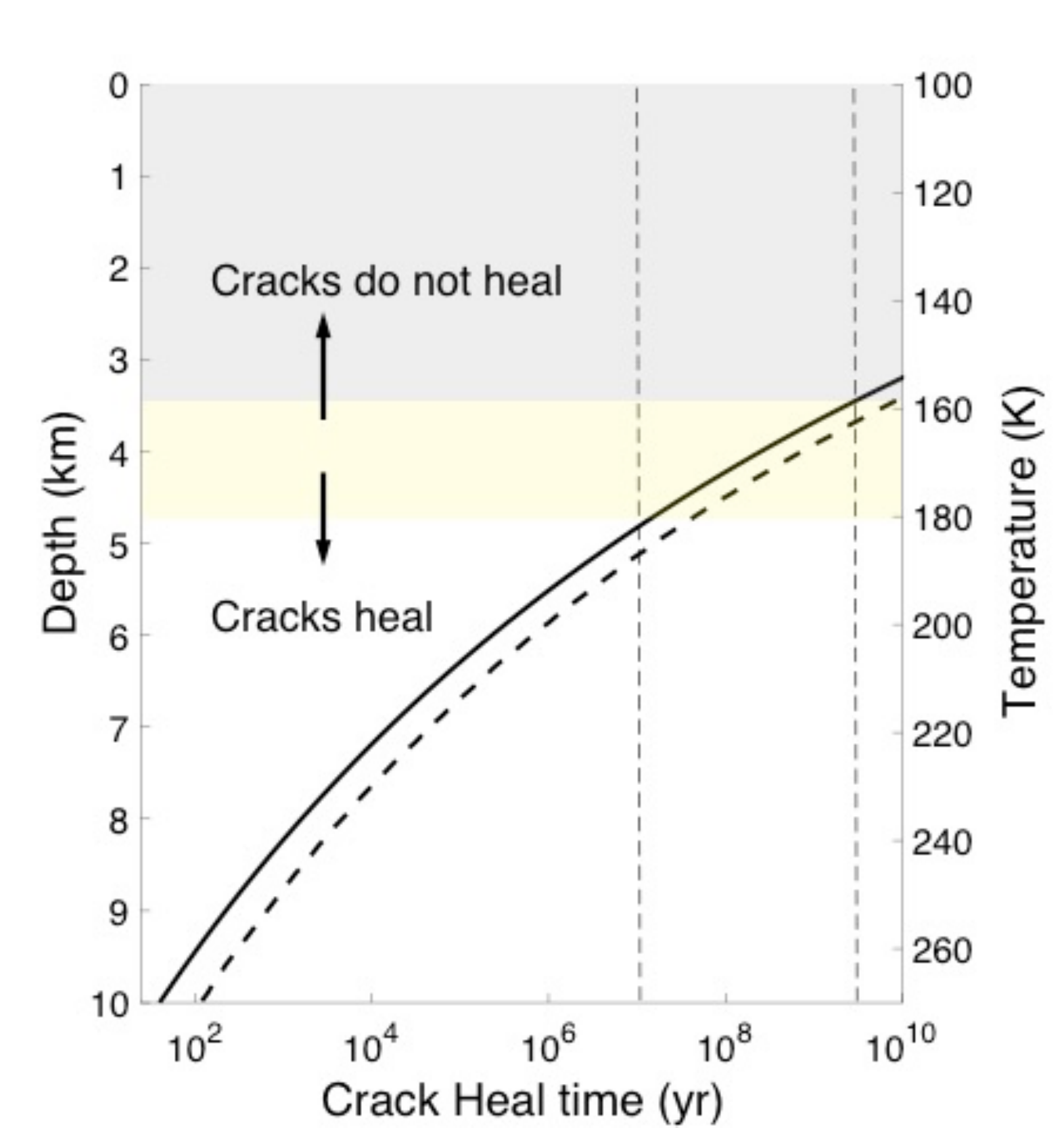}
\caption[Crack healing rates with depth in the ice shell]{The x-axis shows the time scale to heal a $1$ m long crack via diffusion accommodated crack tip retreat. Lines for crack tip radius $r=0.01$ mm and $r=0.1$ mm are shown as solid and dashed lines, respectively. The y-axis shows temperature and approximate depth in the ice shell of Europa assuming the ice shell is conductive with a thermal gradient of $\sim1.7$ K/km. Vertical thin dashed lines show the approximate surface age of Europa and the age of the solar system. At temperature below $160$ K, the time scale for crack healing is greater than $10^9$ years. \label{fig:Figure7}}
\end{figure} 

\end{document}